\documentstyle[amssymb,aps,prb,twocolumn]{revtex}

\begin{document}
\title{Magnetic irreversibility and relaxation in assembly of ferromagnetic
nanoparticles}
\author{R. Prozorov}
\address{Loomis Laboratory of Physics, University of Illinois at Urbana-Champaign,\\
1110 W. Green St., Urbana, IL 61801, U.S.A.}
\author{Y. Yeshurun}
\address{Institute of Superconductivity, Department of Physics, Bar-Ilan University,\\
52900 Ramat-Gan, Israel}
\author{T. Prozorov, A. Gedanken}
\address{Department of Chemistry, Bar-Ilan University, 52900 Ramat-Gan, Israel}
\date{July 21, 1998}
\maketitle

\begin{abstract}
Measurements of the magnetic irreversibility line and time-logarithmic decay
of the magnetization are described for three $Fe_{2}O_{3}$ samples composed
of regular amorphous, acicular amorphous and crystalline nanoparticles. The
relaxation rate is the largest and the irreversibility temperature is the
lowest for the regular amorphous nanoparticles. The crystalline material
exhibits the lowest relaxation rate and the largest irreversibility
temperature. We develop a phenomenological model to explain the details of
the experimental results. The main new aspect of the model is the dependence
of the barrier for magnetic relaxation on the instantaneous magnetization
and therefore on time. The time dependent barrier yields a natural
explanation to the time-logarithmic decay of the magnetization. Interactions
between particles as well as shape and crystalline magnetic anisotropies
define a new energy scale that controls the magnetic irreversibility.
Introducing this energy scale yields a self-consistent explanation of the
experimental data.
\end{abstract}


\section{Introduction}

A ferromagnetic particle becomes monodomain if its size $d$ is reduced below
a critical value $d_{cr}\sim 1-100\ nm$, determined by the competition
between dipole and exchange energies \cite{frenkel30,Landau}. Below this
critical size, the energy loss due to creation of magnetic domain walls
(proportional to $d^{2}$) is larger than the gain due to disappearance of
the dipole magnetic field energy (proportional to $d^{3}$). Such monodomain
ferromagnetic particles can be viewed as large magnetic units, each having a
magnetic moment of thousands Bohr magnetons. Usually neighboring particles
are well separated ($10-20$ $nm$), and direct exchange between particles may
be neglected. Thus, the magnetic properties of an assembly of nanoparticles
are determined by the dipole field energy along with thermal and magnetic
anisotropy energies (see e.g. \cite{Montgomery31,Elmore38,bean59,Bean56}).

Experiments conducted on magnetic nanoparticles show irreversible magnetic
behavior below the ''irreversibility line'' $T_{irr}\left( H\right) $. In
particular, the zero-field cooled (ZFC) and field cooled (FC) magnetization
curves do not coincide, and magnetic hysteresis appears in $M$ vs. $H$
curves (see e. g. \cite{luo91,hanson95,Mydosh96,Delmuro97}). Moreover,
time-logarithmic magnetic relaxation, towards the thermodynamic equilibrium
state, is observed below $T_{irr}\left( H\right) $. Similar observations are
reported here for three systems of $Fe_{2}O_{3}$ nanoparticles with
different shape and crystalline magnetic anisotropies. These nanoparticle
samples were prepared by a sonochemical method, which produces ''regular''
amorphous nanoparticles \cite{suslick91,Cao97,prozorov98}. Sonochemical
irradiation carried out in the presence of magnetic field results in
synthesis of acicular amorphous nanoparticles \cite{prozorov98}. Annealing
of amorphous particles leads to crystallization. In this manner we have
prepared regular amorphous, acicular amorphous and crystalline ferromagnetic
nanoparticles. This enables a study of the effect of shape and crystalline
anisotropies on magnetic irreversibility and the relaxation rate. We find
that, qualitatively, all three samples exhibit similar irreversible magnetic
behavior. However, their irreversibility lines and relaxation rates differ
significantly.

Irreversible magnetic behavior similar to that described here is also
observed in other systems. A noticeable example is magnetic irreversibility
in superconductors \cite{anderson62,beasley69,yeshurun}. In such systems the
origin for irreversibility is the interplay between thermal energy and some
energy barrier, which prevents magnetic reorganization in those materials.
The microscopic origin of the barrier, however, depends on the system. The
magnetic irreversibility in nanoparticles is conventionally associated with
the energy required for a particle moment reorientation, overcoming a
barrier due to magnetic shape or crystalline anisotropy. It is important to
note that the barrier is considered to be independent of the magnetic moment
itself \cite
{hanson95,Delmuro97,Morup94,Coffey96a,Wegrowe97,sanchez95,aharony,Morup83,Morup94a,Bocquet92a,elhilo92,bitoh96,Coffey96,Balcells97,Delmuro97a,Raikher97,Jen91,Grinstaff93}%
. In superconductors, magnetic irreversibility is due to the inevitable
spatial fluctuation of the superconducting order parameter caused by
defects, imperfections etc.; the barrier is the energy required to overcome
the pinning due to this disorder.

An important concept in the theory of irreversible magnetic properties of
superconductors, based on the work of Anderson \cite
{anderson62,beasley69,yeshurun}, is that the effective barrier for magnetic
relaxation increases with time. This is because the superconducting
shielding current (proportional to the magnetization) decays with time,
causing a decrease in the Lorentz force which drives the fluxons away of
their positions. In a nice paper Lottis {\it et al}. \cite{lottis} have put
forward similar arguments to study slow dynamics. They noticed the close
analogy between ferromagnetic assemblies and superconductors. Analyzing the
results of numerical computations they concluded that the decay of the
demagnetizing field is the origin of what the ''quasi-logarithmic''
relaxation. Although the distribution of particle sizes may explain
quasi-time-logarithmic relaxation in a limited time-interval, it is not
necessary for the explanation of the experimentally observed
time-logarithmic relaxation. This approach was later employed in other
studies of the magnetic relaxation, for example, in thin magnetic films \cite
{kirby}.

In this work we adopt the concept of time varying barrier and derive the
phenomenological model to explain magnetic irreversibility and logarithmic
magnetic relaxation in nanoparticles. The physics for the time dependence of
the barrier in nanoparticles is related to the fact that the effective
barrier for reorientation of the magnetic moment of each nanoparticle
depends on the internal magnetic field, which includes the average dipole
field from surrounding nanoparticles. This average dipole field decreases
with time due to the increase of randomness in the orientation of the
magnetic moments of the surrounding nanoparticles. This, in turn, causes the
increase of the effective barrier with time, yielding a natural explanation
to the experimental observation of time-logarithmic relaxation, and a sample
dependent irreversibility line and relaxation rate. Interactions between
particles as well as shape and crystalline magnetic anisotropies define a
new energy scale that controls the magnetic irreversibility. Introducing
this energy scale yields a self-consistent explanation of the experimental
data.

This article is organized in the following way. In Section II we describe
the preparation of the three nanoparticle systems. We then describe our
experimental results of irreversible magnetic properties at various
temperatures, fields and times. In Section III we describe our
phenomenological model and derive equations for the irreversibility line and
the magnetic relaxation. In Section IV we compare the predictions of our
model with the experimental results.

\section{Experimental}

\subsection{Sample preparation and characterization}

\strut Three $Fe_{2}O_{3}$ samples composed of regular amorphous, acicular
amorphous and crystalline, nanoparticles were prepared by a sonochemical
method \cite{suslick91,Cao97,prozorov98}. For the ultrasound irradiation we
used $VC-600$ Sonics and Materials sonicator with $Ti$ horn at $20\ kHz$ and 
$100\ W\cdot cm^{-2}$. In Table I we summarize their features. One molar
solution of $Fe(CO)_{5}$ in decaline was sonochemically irradiated for three
hours in ambient pressure at $0\ ^{o}C$. The powder obtained was
centrifuged, washed repeatedly with dry pentane ($6-7$ times, $8500\ rpm$),
and dried in vacuum at room temperature for three hours. The material
obtained has been accumulated from $2-3$ sonications and the total amount of 
$Fe_{2}O_{3}$ was mixed to ensure the reliability of the results. Then, in
order to remove organic residue, material was annealed in vacuum at $%
140-150\ ^{o}C$ for three hours. Heating up to this temperature was
necessary to evaporate residua of solvents, particularly decaline which has
high a boiling point ($189-191\ ^{o}C$). The amorphous nature of the
particles is confirmed by X-ray diffraction, differential scanning
calorimetry (DSC) analysis and electron-diffraction patterns at selected
areas as shown in Figs. $1a$, $2a$, and the inset to $3a$, respectively. The
absence of Bragg peaks in Fig. $1a$ demonstrates the absence of the
long-range order in the atomic structure; the large endothermic peak in Fig. 
$2a$ indicates an amorphous to crystalline transition at $\thicksim 400$ $%
^{o}C$. The electron diffraction pattern of the inset to Fig. $3a$ also
confirm the amorphous nature of the particles. A typical particle size of $%
\thicksim 50\ nm$ is inferred from the transmission electron micrography
(TEM) picture of Fig. $3a$.

Acicular amorphous particles have been prepared by performing sonication in
external magnetic field of $7$ $kG$ for three hours. The sonication has been
carried out in the $0.25M$ solution of $Fe(CO)_{5}$ in a flask open to air.
We then repeat the wash and dry procedure as described above. The amorphous
nature of the particles was confirmed by X-ray diffraction, DSC, and
selected area electron-diffraction patterns as shown in Figs. $1b$, $2b$,
and the inset to $3b$, respectively. A typical particle length of $\thicksim
50$ $nm$ and diameter of $\thicksim 5$ $nm$ are inferred from TEM picture of
Fig. $3b$.

Heating of amorphous $Fe_{2}O_{3}$ up to $370-380$ $^{o}C$ in ambient
atmosphere for $3-4$ hours resulted in {\it crystalline} $\gamma
-Fe_{2}O_{3} $ nanoparticles. The nature and the internal structure of the
crystalline iron oxide were determined using the X-ray diffraction shown in
Fig. $1c$. The DSC data, Fig. $2c$, do not show any endothermic peak. The
TEM image of Fig. $3c$ show particles of mean size of $\thicksim 200\ nm$.

The second column in Table I summarizes typical particle size for the three
samples. The third column includes the total surface area of the particles,
as measured by BET absorption using $N_{2}$ gas as absorbent.

\subsection{Magnetic measurements procedure}

A {\it Quantum Design} MPMS SQUID magnetometer was used for all magnetic
measurements reported here. The irreversibility line was determined from ZFC
and FC magnetization measurements. Before taking a data point temperature
was stabilized with $0.05\ K$ accuracy and a $30\ \sec $ pause was
sustained. The temperature at which ZFC and FC merge for a constant field $H$
is defined as irreversibility temperature $T_{irr}(H)$. We define the
merging point using a criterion $\left| M_{FC}-M_{ZFC}\right| \approx 0.1\
emu/g.$

The procedure for measurements of magnetic relaxation at different
temperatures is as follows: The sample is cooled in $H=2$ $Tesla$ from a
room temperature (larger than $T_{irr}(2\ Tesla)$) to a target temperature $%
T $, the magnetic field is then reduced to $500\ G$ and the magnetic moment
is measured for approximately two hours. The first data point is taken
approximately two minutes after the field change.

The field dependence of the magnetic relaxation rate is measured at $T=20\ K$%
. At this temperature the field is ramped up to $H=2$ $Tesla$ and reduced
back to a target field $H$, from where the measurements start. The same has
been repeated for negative field $H=-2$ $Tesla$ with consequent a increase
of the magnetic field to a target value and measurements of the magnetic
relaxation.

\subsection{Results}

The experimental results in this section are organized as follows: we first
show $M\left( T,H=\text{constant}\right) $ data, and related measurements of
magnetic relaxation at different temperatures. From the merging point of the
ZFC and FC magnetization curves we extract the irreversibility line for the
three samples. From the relaxation measurements we deduce the relaxation
rate, as a function of temperature, for the three samples. We then present
measurements of magnetization loops $M\left( T=\text{constant},H\right) $
and magnetic relaxation at different values of external field. The
relaxation rate, as a function of field, is then deduced for the three
samples.

Fig. $4$ exhibits typical results of ZFC-FC magnetization curves and
magnetic relaxation at $500\ G$ for the sample composed of amorphous round
nanoparticles. The vertical lines of open circles in Fig. $4$ depict the
relaxation measurements at different temperatures. The vertical arrow
indicates the direction of the time increase. The magnetic moment relaxes
towards the equilibrium moment $M_{rev}$, determined by the FC curve. In the
inset to Fig. $4$ we zoom out at the ZFC-FC curves and indicate by an arrow,
the experimental definition of $T_{irr}$.

The magnetic relaxation data of Fig. $4$ are re-plotted in Fig. $5$ as a
function of time. The solid lines in Fig. $5$ are linear fits for $M$ $%
\propto \ln \left( t\right) $. A qualitatively similar time-logarithmic
decay is also observed in the other two samples. Quantitative differences
will be discussed below.

We define the ''normalized relaxation rate'' $R=\left| \partial M/\partial
\ln \left( t\right) \right| /M_{c}$, i. e., the logarithmic slope of the
relaxation curve normalized by the magnitude of the irreversible
magnetization at which the relaxation starts, $M_{c}=M_{0}-M_{rev}$. Here $%
M_{0}$ is the initial value of the total magnetic moment and $M_{rev}$ is
the magnetic moment corresponding to a field cooling in $500$ $G$. Fig. $6$
summarizes the values of $R$ as a function of temperature, for the three
samples. At low enough temperatures, $R$ is the lowest for the crystalline
sample, intermediate for the acicular amorphous sample and the largest for
the regular amorphous sample. Note, that at higher temperatures it looks as
if $R\left( T\right) $ curves will cross. This is due to a large difference
in the absolute values of $T_{irr}$ ($90$ $K$, $162\ K$ and $216\ K$ at $%
500\ G$ for regular amorphous, acicular amorphous and crystalline,
respectively). As shown in Fig. $16$, $M\left( T\right) $ curves scale with $%
T_{irr}$ and, therefore, in the inset to Fig. $6$ we plot $R$ vs. $T/T_{irr}$%
. In this presentation, the whole $R\left( T/T_{irr}\right) $ curve of the
crystalline sample is lower than that of the acicular amorphous sample and
both are lower than the $R$ curve of the regular amorphous sample.

In Fig. $7$ we compare the irreversibility lines for the three samples. The
largest irreversibility is found in a crystalline sample, intermediate in
the sample with acicular particles and the lowest in the regular amorphous
sample. We explain these observations below.

Magnetic irreversibility below $T_{irr}$ is also demonstrated by measuring
the magnetization loops $M\left( H\right) $. As an example, we show in Fig. $%
8$ $M\left( H\right) $ for the amorphous nanoparticles at $T=5$ and $100\ K$%
. Magnetic hysteresis is apparent at $5\ K$, whereas the behavior is purely
reversible at $100\ K$.

The relaxation at different values of the external magnetic field is shown
in Fig. $9$. The vertical lines represent $M\left( t\right) $ curves shown
along with the standard magnetization loop. The field dependencies of the
relaxation rates for our samples are shown in Fig.$10$. There is an apparent
change in $R$ between low and high fields. At lower fields $R$ is the
largest in an amorphous sample, whereas at large fields the relaxation rate
in an amorphous sample is the lowest.

\section{Magnetic relaxation in the assembly of nanoparticles}

\subsection{Time dependent effective barrier for magnetic reorganization}

Magnetic relaxation is a distinct feature of systems with interacting
particles, far from thermodynamic equilibrium. In an assembly of
ferromagnetic nanoparticles, the elementary process of a change in the
magnetization is the rotation of the magnetic moment of a nanoparticle (or
cluster of such magnetic moments). In the following we assume that the
magnetic anisotropy of each nanoparticle is strong enough to utilize an
Ising-like model, i. e., the magnetic moment of each particle is aligned
only along the anisotropy axis. In Fig. $11a$ we illustrate schematically
the orientation of the elementary magnetic moments of several of such
nanoparticles. The full arrows represent the size and direction of each
magnetic moment. The experimentally measured magnetic moment is determined
by the sum of the projections of each individual particle's moment on the
direction of the external magnetic field. Note that the directions of the
easy axes are randomly distributed. For such a system, the energy $W$ of
each magnetic nanoparticle, neglecting for the moment the interparticle
interactions, varies with the angle as \cite{aharony}:

\begin{equation}
W=KV\sin ^{2}\left( \varphi -\theta \right) -M_{p}H\cos \left( \varphi
\right)  \label{Wsimple}
\end{equation}
Here $\theta $ is the angle between the easy axis $\overrightarrow{K}$ and
the external magnetic field $\overrightarrow{H}$, and $\varphi $ is the
angle between the particle magnetic moment $\overrightarrow{M}_{p}$ and the
external field. In order to have any magnetic irreversibility and
relaxation, the $KV$ term in Eq. \ref{Wsimple} must be larger than the $%
M_{p}H$ term and we will consider this limiting case. The reduced energy $%
W/KV$ of Eq. \ref{Wsimple} is plotted in Fig. $12$ as a function of the
angle $\varphi $ for two different fields $H_{1}=2.5KV/M_{p}$ (bold) and $%
H_{2}=0.5KV/M_{p}$ (light). Since the magnetic anisotropy has no preferable
direction, there are two minima in the angular dependence of the energy, as
shown in Fig. $12$. The external magnetic field fixes the direction of the
lowest minima. We denote by $U_{12}$ the barrier for reorientation from the
lowest minima ($W_{1}$) to the other minima ($W_{2}$). The backward
reorientation requires overcoming the energy barrier $U_{21}.$

In order to take inter-particle interactions into account we view the field $%
H$ in Eq. \ref{Wsimple} as the internal magnetic field, which is the sum of
the external field and the dipole field from the surrounding nanoparticles.
This local magnetic field depends on the directions of neighboring magnetic
moments \cite{bouchard93,Zaluska93}. Since the magnetic moment is a
statistical average of those moments, the local field depends, on the
average, on the total magnetic moment. This induces a feedback mechanism:
each reorientation of an individual nanoparticle decreases the total
magnetic moment. This is illustrated in Figs. $11a$. and $11b$. Figure $11a$
represents a snapshot of a field-cooled system of nanoparticles in which
most of the individual magnetic moments are favorably oriented in a
direction such that their projections are along the external field. After a
field decrease, as a result of thermal fluctuations, some magnetic moments
reorient so that their projection is anti-parallel to the external field.
The open arrows in Fig. $11b$ represent those reoriented moments.

Since the local dipole field decreases during this process, the average
barrier $U_{12}$ increases. As indicated in the Introduction, an increase of
the barrier with time is a characteristic of other irreversible systems,
such as type-II superconductors in the process of magnetic flux creep.

The dynamics resulting from such a scenario is sketched Fig. $13$.
Immediately after reducing the magnetic field, individual magnetic moments
are still along the direction of the external field, i. e., in minima $W_{1}$
of Fig. $13$, as depicted by the population of the black dots. During the
relaxation process magnetic moments flip to the minimum $W_{2}$ in the
figure. Since, as discussed above, this barrier depends on the total
magnetic moment via dipole fields, it will increase with time as shown in
the figure, with dipole fields working on the average against the external
field. The total magnetic moment along the magnetic field is thus decreased,
as sketched in Fig. $11b$.

\subsection{Equations of magnetic relaxation}

In a realistic sample, the directions of easy axes are randomly distributed,
the particles cannot physically rotate (e. g., in a dense powder of
ferromagnetic nanoparticles), and dipole interactions are strong. We will
model this situation as outlined below.

Any given particle $i$ in Fig. $11$ has an anisotropy axis at a fixed angle $%
\theta _{i}$ relative to the external magnetic field. The magnetic moment of
this particle is then oriented at an angle $\varphi _{i}$ to the field. This
angle is defined by the non-local energy minimization, due to dipole fields
of the surrounding. It is important to note that each particle interacts
with a local magnetic field $H_{i}$ which is the result of a vector sum of
the external and dipole fields. At small enough external field and large
enough anisotropy $\varphi _{i}$ may have two values: $\varphi _{i}\approx
\theta _{i}$ or $\varphi _{i}\approx $ $\theta _{i}+\pi $, which leads to
the situation described in Fig. $12$, with two energy minima at $%
W_{1}^{i}\approx -M_{p}H_{i}\cos \left( \theta _{i}\right) $ and $%
W_{2}^{i}\approx M_{p}H_{i}\cos \left( \theta _{i}\right) $. Thermal
fluctuations may force particle moment in the minima $W_{1}^{i}$ to change
its direction to another minima $W_{2}^{i}$ and vise versa. The $%
W_{1}^{i}\rightarrow W_{2}^{i}$ rotation requires overcoming a barrier $%
U_{12}^{i}$, and a barrier $U_{12}^{i}$ for backward rotation, see Eqs. \ref
{U21} and \ref{Us12} of the Appendix, respectively. We then assume that the
field $H_{i}$ can be represented as a simple sum of the external field $H$
and the collinear to $H$ dipole field $H_{d}$ (i.e. independent of $\theta
_{i}$). The amplitude of a dipole field $H_{d}$ at any given site depends
upon orientations of the moments of the surrounding particles. If those
orientations are totally random (minima $W_{1}$ and $W_{2}$ are equally
occupied) the dipole field is small, whereas if all surrounding particles
are in one of the minima the resulting dipole field is maximal. From this
simple analysis, we conclude that the magnitude of a dipole field depends
upon the total magnetic moment $M$.

Considering the balance of forward and backward rotations, and averaging
over the volume of the sample, we show in the Appendix that magnetic
relaxation is described by a differential equation similar to that derived
for superconductors \cite{beasley69,yeshurun}:

\begin{equation}
\frac{\partial M}{\partial t}=-AM_{c}\exp \left( -\frac{U}{T}\right)
\label{dMdtMc}
\end{equation}
where $A$ is an attempt frequency and $U$ is an effective barrier for
magnetic relaxation given by

\begin{equation}
U=U_{0}\left( 1-\frac{M}{M_{0}}\right)  \label{U(M)}
\end{equation}
where

\begin{equation}
U_{0}=2KV+4M_{p}\left( H-\gamma M_{rev}\right) /\pi  \label{U0}
\end{equation}
and $M_{0}=\frac{1}{\gamma }\left( \frac{\pi KV}{2M_{p}}+H-\gamma
M_{rev}\right) $. Here $\gamma $ is the constant accounting for the strength
of the dipole-dipole interactions, $M_{p}$ is the magnetic moment of an
individual particle, $K$ is the anisotropy constant and $V$ is the particle
volume. Apparently, as $\gamma \rightarrow 0$ the energy barrier $%
U\rightarrow U_{0}$, thus $U_{0}$ is the barrier in the assembly of
non-interacting particles. It is worth noting that in our model the barrier $%
U$ depends on the magnetic moment in the same way as that used by Anderson 
\cite{anderson62,beasley69,yeshurun} for a description of magnetic
relaxation in superconductors.

In the following we analyze magnetic relaxation described by those equations.

If the barrier for a particle moment reorientation does not depend on the
total magnetic moment, i.e., $\gamma =0$ and $U=U_{0}$, direct integration
of Eq. \ref{dMdtMc} yields: 
\begin{equation}
M=M_{c}\exp \left( -t/\tau \right)  \label{Mnon}
\end{equation}
where $M_{c}$ is the initial irreversible magnetization and $\tau =\exp
\left( U_{0}/T\right) /A$ is the {\em macroscopic} characteristic relaxation
time. This result is very similar to that derived in early works for
classical N\'{e}el's superparamagnetic relaxation, see e.g., \cite{neel49}.
This exponential decay is observed experimentally, for example in the work
of Wegrowe {\it et al.} \cite{Wegrowe97} on a single nano-wire.

If interactions are not negligible, Eq. \ref{dMdtMc} may be rewritten in
dimensionless form: 
\begin{equation}
\frac{\partial u}{\partial \widetilde{\tau }}=-\exp \left( -u\right)
\label{dudt}
\end{equation}
where $u=U/T$ and $\widetilde{\tau }=t/\widetilde{t}$ with 
\begin{equation}
\widetilde{t}=\frac{M_{0}}{M_{c}}\frac{T}{U_{0}}\frac{1}{A}=\frac{\pi T}{%
4\gamma AM_{p}M_{c}}=\frac{T}{A\Theta }  \label{ttilda}
\end{equation}
where we introduced a new energy scale $\Theta $, which, as we show below,
determines the relaxation process and the irreversibility line: 
\begin{equation}
\Theta =M_{c}U_{0}/M_{0}=4\gamma M_{c}M_{p}/\pi  \label{theta}
\end{equation}
This energy is directly related to the strength of the interparticle
interactions.

Solving Eq. \ref{dudt} we obtain 
\begin{equation}
u=u_{c}+\ln \left( 1+\frac{t}{t_{0}}\right)  \label{U}
\end{equation}
where $u_{c}=U_{c}/T$ is the reduced effective energy barrier at $t=0$, the
time when the relaxation starts. The normalization time $t_{0}$ is given by 
\begin{equation}
t_{0}=\widetilde{t}\exp \left( \frac{U_{c}}{T}\right) =\frac{T}{A\Theta }%
\exp \left( \frac{U_{0}-\Theta }{T}\right)  \label{t0}
\end{equation}
Now, using Eqs. \ref{U(M)} and \ref{U} we get the time evolution of the
magnetic moment: 
\begin{equation}
M\left( t\right) =M_{c}\left( 1-\frac{T}{\Theta }\ln \left( 1+\frac{t}{t_{0}}%
\right) \right)  \label{M(t)}
\end{equation}

Normalized relaxation rate $R\equiv \left| \partial M/\partial \ln \left(
t\right) \right| /M_{c}$ is given by: 
\begin{equation}
R=\frac{T}{\Theta }\frac{t}{t_{0}+t}  \label{R(t)}
\end{equation}
As we will see below, experiment shows that $t_{0}<1$ sec. In our
measurements typical time window $\Delta t\approx 100$ sec, therefore we can
assume $t\gg t_{0}$ and Eq. \ref{R(t)} predicts that the relaxation rate
saturates at $R=T/\Theta $. Thus, measurements of the normalized relaxation
rate can provide direct estimate of the energy scale $\Theta $ governing the
relaxation process.

\subsection{Irreversibility temperature}

The irreversibility temperature $T_{irr}$ of the assembly of magnetic
nanoparticles is defined by the condition $M\left( \Delta t,T_{irr}\right)
=\Delta M.$ Here $\Delta M$ is the smallest measured magnetic moment and $%
\Delta t$ is the time window of the experiment. Using Eq. \ref{M(t)} we
obtain:

\begin{equation}
T_{irr}=\Theta \frac{1-\Delta M/M_{c}}{\ln \left( 1+\Delta t/t_{0}\right) }%
\approx \frac{\Theta }{\ln \left( 1+\Delta t/t_{0}\right) }  \label{Tirr}
\end{equation}
thus we can estimate the characteristic time $t_{0}$ from measurements of $%
T_{irr}$, because the energy $\Theta $ can be determined separately from the
measurements of the relaxation rate $R\approx T/\Theta $. On the other hand
the irreversibility line $T_{irr}\left( H\right) $ gives the field
dependence of $\Theta $. The latter may be obtained also from $R\left(
H\right) $ measurements. Thus measurements of $T_{irr}\left( H\right) $ and $%
R\left( H\right) $ in different samples provide a verification of our model
on self-consistency.

It is interesting to note that the expression for $T_{irr}$, Eq. \ref{Tirr},
is typical for the blocking temperature of individual non-interacting
particles, which is obtained from Eq. \ref{Mnon}:

\begin{equation}
T_{irr}^{0}=\frac{U_{0}}{\ln \left( \Delta t/t^{\ast }\right) }
\label{tirr0}
\end{equation}
where $t^{\ast }=1/\left( A\ln \left| M_{c}/\Delta M\right| \right) $ is the
characteristic time and $U_{0}$ is given by Eq. \ref{U0}. Energy $U_{0}$ is
proportional to $KV$ for non-interacting nanoparticles, but it is reduced by
a term proportional to $\gamma $ due to interparticle interactions. This is
in agreement with previous works where ''static'' modifications of the
barrier for relaxation were considered \cite{Morup94,Morup94a,Morup93}.
Irreversibility temperature, Eq. \ref{tirr0} approaches $0$ when $\Delta
M\rightarrow 0$, and so does $T_{irr}^{0}$. This reveals an important
difference in the physics of the irreversibility line in interacting and
non-interacting particles. In the former, there is a true irreversibility in
the limit $\Delta M\rightarrow 0$ associated with freezing of magnetic
moments due inter-particle interactions. In the case of non-interacting
nanoparticles, the apparent irreversibility is due to experimental
limitations (finite sensitivity, e.g., $\Delta M$). It is important to
stress that this is true only on a macroscopic time scale $\Delta t\gg
t^{\ast }$, such as relaxation or $M\left( T\right) $ measurements. If,
however, $\Delta t<t^{\ast }$ is realized, for example in M\"{o}ssbauer
measurements, one may detect the irreversibility temperature according to
Eq. \ref{tirr0} \cite{Morup94,Bocquet92a,linderoth93}.

We also note that in any case $T_{irr}$ is a dynamic crossover from
reversible to irreversible state and is defined for a particular
experimental time window $\Delta t$.

In the following section we compare our experimental observations with the
model developed above.

\section{ Discussion}

Our phenomenological model provides a description of the irreversible
magnetic behavior in the assembly of ferromagnetic nanoparticles. In
particular, the model predicts the time-logarithmic decay of the
magnetization, see Eq. \ref{M(t)}. Also, Eqs. \ref{R(t)} and \ref{Tirr}
relate both the irreversibility line and the relaxation rate to a single
parameter $\Theta =4\gamma M_{c}M_{p}/\pi $.

The magnetic relaxation data of Fig. $5$ reveal, indeed, time-logarithmic
relaxation. Fitting these data to Eq. \ref{M(t)} yields the parameters of $%
M_{c}$ and $\Theta $. In Fig. $14$ we plot the derived energy $\Theta $ as a
function of temperature for the three samples and find that $\Theta $ is the
largest for a crystalline sample, intermediate for an acicular amorphous and
the lowest for a regular amorphous sample. The straightforward explanation
is that in a crystalline sample both $\gamma $ and $M_{p}$ are the largest;
in an acicular amorphous sample $M_{p}$ is of the same order as in regular
amorphous, but $\gamma $ is much larger due to shape anisotropy.

Similarly, we derive the magnetic field dependence of $\Theta $ from the
data of Fig. $9$ and plot $\Theta \left( H\right) $ for three samples in
Fig. $15$. We note that $M_{p}$ and $\gamma $ should not depend on magnetic
field. It is therefore expected that the field dependence of $\Theta $ is
determined by the field dependence of $M_{c}\approx $ $M_{s}-M_{rev}(H)$,
which decreases with field, Fig. $9$. Figure $15$ shows the agreement with
this observation. The weak increase of $\Theta $ with temperature, Fig. $14$%
, may be related to some non-linear dependence of barrier $U$ on the
magnetic moment.

Independent estimations of $\Theta $ are derived from $T_{irr}$ of Fig. $7$
using Eq. \ref{Tirr}. Comparing Fig. $7$ and Fig. $15$ we get (for three
samples) $\Theta /T_{irr}\approx 4-6$. Thus, $t_{0}\approx 0.05-0.5\ \sec $.
Note that these values of $t_{0}$ are much larger than the ''microscopic''
values predicted by N\'{e}el \cite{neel49}, simply because they reflect
collective behavior of the whole assembly controlled by the effective
barrier $\Theta $, see Eq.\ref{t0}, and not a single particle barrier $KV$.

Let us now compare the irreversibility lines of different samples, Fig. $7$.
In most parts of this diagram the region of the irreversible behavior is the
largest for a crystalline sample. The amorphous sample containing acicular
particles occupies the intermediate space and the amorphous sample embraces
the smallest space in this $T-H$ phase diagram. Such behavior is naturally
explained in terms of a strength of inter-particle interactions, which are
the smallest in the case of a regular amorphous sample, intermediate for an
acicular amorphous sample (due to shape anisotropy) and the largest for a
crystalline sample due to crystalline anisotropy. Also, highest
irreversibility temperature of crystalline sample is understood on the basis
of its largest particle size.

\section{Summary and conclusions}

We presented measurements of irreversible magnetization as a function of
temperature, time and magnetic field in three types of ferromagnetic
nanoparticles: regular amorphous, acicular amorphous and crystalline
nanoparticles. The results are interpreted using a developed
phenomenological approach based on the assumption that the barrier for
magnetic moment reorientation depends on the total magnetic moment via
dipole fields. This explains the time-logarithmic magnetic relaxation
governed by the energy scale $\Theta $ related to interparticle interaction.
Values of $\Theta $ found from measurements of the irreversibility line and
the relaxation rate are in perfect agreement, implying validity of our model.

{\em Acknowledgments:} We thank Y. Rabin and I. Kanter for valuable
discussions. This work was partially supported by The Israel Science
Foundations and the Heinrich Hertz Minerva Center for High Temperature
Superconductivity. Y.Y. acknowledges support from the German Israeli
Foundation (G.I.F). R. P. acknowledges support from the Clore Foundations.

\newpage{}

\section{Appendix: Effective barrier for magnetic relaxation and equation
for time evolution of the magnetic moment}

Here we consider in detail the model outlined in the text. We assume that
magnetic moment $M_{p}$ of any given particle $i$ can be in one of the two
possible energy minima: $W_{1}^{i}\approx -M_{p}H_{i}\cos \left( \theta
_{i}\right) $ or $W_{2}^{i}\approx M_{p}H_{i}\cos \left( \theta _{i}\right) $%
. These minima are separated by the barrier of height $\symbol{126}%
KV+M_{p}H_{i}\sin \left( \theta _{i}\right) $. In the presence of thermal
fluctuations, a particle moment sitting in the minima $W_{1}^{i}$ can
spontaneously change its direction to the next minima $W_{2}^{i}$. The
energy barrier for such reorientation is 
\begin{equation}
U_{12}^{i}=KV+M_{p}H_{i}\left( \sin \left( \theta _{i}\right) +\cos \left(
\theta _{i}\right) \right)  \label{U12}
\end{equation}
The backward rotation is also possible and requires overcoming the barrier: 
\begin{equation}
U_{21}^{i}=KV+M_{p}H_{i}\left( \sin \left( \theta _{i}\right) -\cos \left(
\theta _{i}\right) \right)  \label{U21}
\end{equation}

From this point on one can conduct a self-consistent statistical average
over angles $\varphi _{i}\left( H_{i},t,\theta _{i}\right) $ in order to
evaluate the resulting magnetic moment $M$ of the system. On the other hand
we may try to simplify the problem assuming that the internal field $H_{i}$
can be represented as a simple sum of the external field $H$ and the
collinear to it dipole field $H_{d}$ (i.e. independent of $\theta _{i}$). If
all easy axes are randomly distributed the average barrier for flux
reorientation is then given by 
\[
U_{k}\equiv \left\langle U_{i}\right\rangle _{k}=\frac{1}{N_{k}}%
\sum_{i=1}^{N_{k}}U_{i}\approx \frac{2}{\pi }\int\limits_{0}^{\pi /2}U\left(
\theta \right) d\theta 
\]
where $k=12$ or $21$ denotes particle's moment flipping from the minima $%
W_{1}$ to the minima $W_{2}$, or backward, respectively. Using Eq. \ref{U12}
and Eq. \ref{U21} we find

\begin{eqnarray}
U_{12} &=&KV+4M_{p}H_{i}/\pi  \label{Us12} \\
U_{21} &=&KV  \label{Us21}
\end{eqnarray}

Let us now consider a situation where temperature is higher than
irreversibility temperature and system is at thermal equilibrium. The number
of particles jumping per unit time from one minima to another is
proportional to $N_{k}\exp (-U_{k}/T)$. The condition for equilibrium is 
\[
N_{1}e^{-\frac{U_{12}}{T}}=N_{2}e^{-\frac{U_{21}}{T}}
\]
Thus 
\[
N_{2}=N_{1}\exp \left( -\frac{U_{12}-U_{21}}{T}\right) =N_{1}\exp \left( -%
\frac{4M_{p}H_{i}}{\pi T}\right) 
\]
It is clear that the difference $n=N_{1}-N_{2}$ determines the resulting
magnetic moment of a system. If the total number of particles in the system
is $N$ difference, $n$ is 
\begin{equation}
n=N_{1}-N_{2}=N\frac{1-\exp \left( -\frac{4M_{p}H_{i}}{\pi T}\right) }{%
1+\exp \left( -\frac{4M_{p}H_{i}}{\pi T}\right) }=N\tanh \left( \frac{%
2M_{p}H_{i}}{\pi T}\right)   \label{n}
\end{equation}
The total reversible magnetic moment then is 
\begin{equation}
M_{rev}\approx M_{p}n=M_{p}N\tanh \left( \frac{2M_{p}H_{i}}{\pi T}\right) 
\label{Mrev}
\end{equation}
This formula is similar to the expression for the Ising superparamagnet and
simply reflects the two-state nature of our model\cite{neel49,viitala97}.
The difference is, however, that the physical magnetic field is the total
(external + dipole) field $H_{i}$.

Dipole field $H_{d}$ at any given site depends upon orientations of the
moments of the surrounding particles. If those orientations are totally
random (minima $W_{1}$ and $W_{2}$ are equally occupied) the dipole field is
small, whereas if all surrounding particles are situated in one of the
minima the resulting dipole field is maximal. From this simple picture, we
conclude that the magnitude of a dipole field depends upon the total
magnetic moment of a sample $M_{rev}+M$, where $M_{rev}$ is given by Eq. \ref
{Mrev} and $M$ is the irreversible, time dependent contribution to the total
magnetic moment resulting from the finite relaxation time needed for a
system to equilibrate. Therefore, we may write $H_{i}=H-\gamma \left(
M_{rev}+M\right) $. Here $\gamma $ is the coefficient accounting for the
contribution of dipole fields. Now we can obtain the equation for reversible
magnetization from Eq.\ref{Mrev}:

\begin{equation}
M_{rev}\approx M_{S}\frac{\tanh \left( \frac{2M_{p}H}{\pi T}\right) }{%
1+\gamma \frac{2M_{p}M_{S}}{\pi T}}  \label{Mreva}
\end{equation}
where $M_{S}=M_{p}N$. We note that this formula is valid at small enough
fields $2M_{p}H/\pi <T$ when particle moments are almost locked along the
easy axes and small enough interactions (i.e. $H>\gamma M_{rev}$) . The
important result is that reversible magnetization decreases as the
inter-particles interaction increases. Interestingly, Eq. \ref{Mreva}
provides a good description of the experimental data.

Thus, the barriers for moment reorientation in Eq. \ref{Us12} and Eq. \ref
{Us21} can be re-written as 
\begin{eqnarray}
U_{1} &=&KV+4M_{p}\left( H-\gamma \left( M_{rev}+M\right) \right) /\pi
\label{U1} \\
U_{2} &=&KV  \label{U2}
\end{eqnarray}

We shall now consider direct and backward moment rotation processes in a non
equilibrium state. As above, we denote by $N_{1}$ and $N_{2}$ number of
moments in energy minima $1$ and $2$, respectively. The total number of
particles in the system is $N=N_{1}+N_{2}$. The magnetic moment is
proportional to the difference $n=N_{1}-N_{2}$. During small time $\delta t$
this difference changes as 
\begin{equation}
\delta n=\left( N_{1}\exp \left( -\frac{U_{1}}{T}\right) -N_{2}\exp \left( -%
\frac{U_{2}}{T}\right) \right) \delta t  \label{dn1}
\end{equation}
Using simple algebra and above relationships between $N_{1}$, $N_{2}$ , $n$,
and $N$, we get 
\begin{equation}
\frac{\delta n}{\delta t}=-\exp \left( \frac{U_{1}+U_{2}}{T}\right) \left(
n\cosh \left( \frac{U_{1}-U_{2}}{T}\right) +N\sinh \left( \frac{U_{1}-U_{2}}{%
T}\right) \right)  \label{dn}
\end{equation}
From this we arrive to a non-linear differential equation governing process
of magnetic relaxation {\it not too close to equilibrium}: 
\begin{equation}
\frac{dM}{dt}\approx -A\exp \left( \frac{U_{1}+U_{2}}{T}\right) \left(
\left( M_{rev}+M\right) \cosh \left( \frac{U_{1}-U_{2}}{T}\right)
+M_{s}\sinh \left( \frac{U_{1}-U_{2}}{T}\right) \right)  \label{dMdtGENERAL}
\end{equation}
where $A$ is a constant measured in $\sec ^{-1}$ and having meaning of
attempt frequency.

Eq. \ref{dMdtGENERAL} can be simplified considering magnetic relaxation not
too close to equilibrium and retaining our assumption that anisotropy
contribution to the magnetic energy is much larger than that of magnetic
field (both conditions are better satisfied at low fields). In this case,
Eq. \ref{dMdtGENERAL} may be approximated in a reduced form:

\begin{equation}
\frac{\partial M}{\partial t}=-AM_{c}\exp \left( -U/T\right)  \label{dMdta}
\end{equation}
where $M_{c}$ is the total magnetic moment at the beginning of relaxation
and $U$ is the effective barrier: 
\begin{equation}
U=2KV+4M_{p}\left( H-\gamma M_{rev}-\gamma M\right) /\pi =U_{0}\left( 1-%
\frac{M}{M_{0}}\right)  \label{U(M)a}
\end{equation}
where $U_{0}=$ $KV+4M_{p}\left( H-\gamma M_{rev}\right) /\pi $ and $M_{0}=%
\frac{1}{\gamma }\left( \frac{\pi KV}{2M_{p}}+H-\gamma M_{rev}\right) $.

We reiterate that Eq. \ref{dMdta} is valid only in the case when the
magnetic anisotropy is large and magnetic moment is far from equilibrium.
Close to equilibrium, one ought to consider Eq. \ref{dMdtGENERAL}.

\-\newpage {}

\section{Figure captions}

\bigskip

Fig.1 X-ray diffraction patterns for (a) regular amorphous, (b) acicular
amorphous and (c) crystalline samples.

Fig.2 Differential scanning colorimetry spectra for (a) regular amorphous,
(b) acicular amorphous and (c) crystalline samples.

Fig.3 Transmission electron micrographs for (a) regular amorphous, (b)
acicular amorphous and (c) crystalline samples.

Fig.4 Typical ZFC - FC curves with superimposed relaxation for the amorphous
sample measured at $500\ G$ at different temperatures. {\it Inset}:
full-range ZFC-FC curve.

Fig.5 Typical relaxation curves measured in the amorphous sample in $500\ G$
at different temperatures.

Fig.6 Normalized logarithmic relaxation rate $R$ for three types of samples
as a function of temperature. {\it Inset:} $R$ as a function of a reduced
temperature $T/T_{irr}$.

Fig.7 Irreversibility lines for three types of samples.

Fig.8 Typical magnetization loops at $T=5\ K$ (open circles) and at $T=100$\
(solid line).

Fig.9 Magnetic relaxation at different values of magnetic field. Vertical
lines are the $M\left( t\right) $ curves superimposed on a regular
magnetization loop measured at the same temperature.

Fig.10 Normalized logarithmic relaxation rate $R$ for three samples as a
function of magnetic field at $T=20\ K$.

Fig.11 Schematic snapshots of magnetic moments distribution in powder sample
at (a) beginning of the relaxation and (b) at latter time

Fig.12 Energy profiles after FC in magnetic high field ($H_{1}$) and after
reduction of the magnetic field, whence the relaxation starts ($H_{2}$).

Fig.13 Energy profiles at the beginning and at the latter stage of the
relaxation. Dots indicate population of magnetic moments of the particular
energy minima.

Fig.14 Temperature dependence of energy $\Theta $ extracted from the
measurements of normalized relaxation rate for the three samples.

Fig.15 Magnetic field dependence of the energy $\Theta $.

Fig. 16 Scaling of the $M\left( T\right) $ FC-ZFC curves with
irreversibility temperature.

\bigskip {\bf \newpage}

{\bf Table I}. Characteristic parameters of the samples

\bigskip 
\begin{tabular}{||c|c|c||}
\hline\hline
{\bf Sample} & $d\ [nm]$ & {\bf Surface area} $[m^{2}/g]$ \\ \hline\hline
\multicolumn{1}{||l|}{regular amorphous} & \symbol{126}$50$ & $148$ \\ \hline
\multicolumn{1}{||l|}{acicular amorphous} & \symbol{126}$5\times 50$ & $164$
\\ \hline
\multicolumn{1}{||l|}{crystalline} & \symbol{126}$200$ & $88$ \\ \hline\hline
\end{tabular}

\bigskip

\end{document}